# Interplay between optical emission and magnetism in the van der Waals magnetic semiconductor CrSBr in the two-dimensional limit.

*Francisco Marques-Moros, Carla Boix-Constant, Samuel Mañas-Valero, Josep Canet-Ferrer,\* and Eugenio Coronado\**

Instituto de Ciencia Molecular (ICMol), Universitat de València, Paterna, Spain.

Keywords: 2D semiconductors, 2D magnets, van der Waals materials, exciton, trion.

ABSTRACT: The Van der Waals semiconductor metamagnet CrSBr offers an ideal platform for studying the interplay between optical and magnetic properties in the two-dimensional limit. Here, we carried out an exhaustive optical characterization of this material by means of temperature and magnetic field dependent photoluminescence (PL) on flakes of different thicknesses down to the monolayer. We found a characteristic emission peak that is quenched upon switching the ferromagnetic layers from an antiparallel to a parallel configuration and exhibits a different temperature dependence from that of the peaks commonly ascribed to excitons. The contribution of this peak to the PL is boosted around 30-40 K, coinciding with the hidden order magnetic transition temperature. Our findings reveal the connection between the optical and magnetic properties via the ionization of magnetic donor vacancies. This behavior enables a useful tool for the optical reading of the magnetic states in atomically thin layers of CrSBr and shows the potential of the design of two-dimensional heterostructures with magnetic and excitonic properties.

**INTRODUCTION**

Excitons in two-dimensional (2D) semiconductors offer an interesting landscape not only in fundamental terms (dynamics of localized, dark, bright and interlayer excitons) but as well for opto-electronic applications (light emitters, opto-valleytronic devices or solar cells, among others), as extensively explored in group VI transition metal-dichalcogenides ($MX_2$, where M = Mo, W and X = S, Se).[1–6] With the recent discovery of 2D magnetic semiconductors, magnetic fields can be employed for tuning the excitonic dynamics and, more interestingly, coupling them to other quasiparticles, as recently illustrated by coupling excitons and magnons in CrSBr, thus paving new avenues in spintronic and magnonic fields.[7–10]

CrSBr is a direct bandgap layered semiconductor composed of antiferromagnetically ($T_N \sim 140$ K) coupled ferromagnetic layers ($T_C \sim 150$ K),[11,12] corresponding the easy, intermediate, and hard magnetic axis to the *b*, *a,* and *c* crystallographic axis (Figure 1A). This A-type metamagnet exhibits a rich field-induced phenomenology since, by the application of moderate fields, the magnetization of the layers can be switched from antiparallel (AP) to parallel (P) configuration via a spin reversal (in bulk, 0.6 T for fields along the easy axis) and reoriented (in bulk, 1T and 2 T for fields along the intermediate and hard axis), thus behaving like a ferromagnet.[13–15]

The fields required for reaching a parallel alignment of the magnetization exhibit a dimensionality dependence, being reduced from 0.6 T in bulk to 0.2 T in the bilayer case for fields applied along the easy axis.[14] More striking, a low-temperature transition has been reported at *ca.* 40 K (the so-called *hidden-order*) by several groups using different techniques,[13,14,16–19] both in bulk and in atomically-thin layers, although its exact origin is still under discussion. While some authors attribute the change in the magnetic behavior to a spin freezing process,[18] other authors present evidence of interaction with magnetic defects, in particular intrinsic S and Br vacancy centers.[17]

Regarding the optical properties of CrSBr, Wilson *et al.* reported the dependence of photoluminescence in few-layer CrSBr flakes on the temperature and attributed the presence of different peaks in the emission spectra to interlayer Coulomb interactions among carriers.[20] This was experimentally confirmed by studying the dependence of the PL on the external magnetic field, which consists of a dramatic redshift in the excitonic emission (in the order of 20 meV) for bilayer and few layer films. Theoretical predictions in this work indicated that the field-induced ferromagnetic-like state is accompanied by electron delocalization.

Afterwards, Klein and co-workers observed a correlation between the emission of optically active defects and the hidden order transition in thick CrSBr flakes (ca. 40 nm).[17] The authors concluded that either S or Br vacancy centers (i.e. $V_S$ or $V_{Br}$) would be the origin or part of the mechanism that drives the hidden order transition. More recently, the same group observed a correlation between intentionally induced defects in mono- to multi-layer CrSBr and their structural, vibrational, and magnetic properties.[21]

However, important questions regarding the electronic-magnetic coupling in CrSBr crystals are still open. Hence, studying different excitonic species in CrSBr flakes and understanding their connection with the magnetic properties will be of paramount importance for improving the coupling among magnetic and semiconductor materials in vdW heterostructures. To this end, we focus in this work on the study of the exciton dynamics in CrSBr down to the single-layer case and its correlation with the magnetism. In general, excitonic emission is reduced by increasing the temperature above the hidden order transition. In contrast, we have found an isolated peak exhibiting the opposite behaviour, an increasing contribution with temperature above the hidden order transition. In this work we will demonstrate that this peak (namely P4) enables the monitoring of the magnetic states of few-layer CrSBr flakes by optical measurements and illustrates the connection between the optical and the magnetic properties via impurities (magnetic donor vacancies in the present case).

**RESULTS AND DISCUSSION**

The crystalline structure of CrSBr consists of vdW layers made of two planes of CrS terminated with Br atoms, see Figure 1A. The samples under study consist of several mechanically exfoliated CrSBr flakes deposited over a silicon substrate into an inert atmosphere. The anisotropic crystal structure led to the generation of elongated flakes by means of mechanical exfoliation. Despite recent works claim the air stability of CrSBr flakes[13,22,23], the monolayer, the bilayer, and trilayer flakes were covered with hBN for further protection during the manipulation of the samples, see Figure 1B.

INFLUENCE OF THE THICKNESS ON THE OPTICAL PROPERTIES

We start our study by comparing the emission of CrSBr flakes of different thicknesses by means of micro-photoluminescence (μ-PL). At low temperatures, the PL is composed of two bands, labeled as B1 and B2 in Figure 1C. In the case of thicker flakes, B1 appears as a broad band centered around 1.23 eV, while this band is absent when approaching the 2D limit. The emission energy of B1 coincides with the energy reported for the emission of impurities in bulk CrSBr, ascribed to Br or S vacancies ($V_{Br}$ or $V_S$).[17]

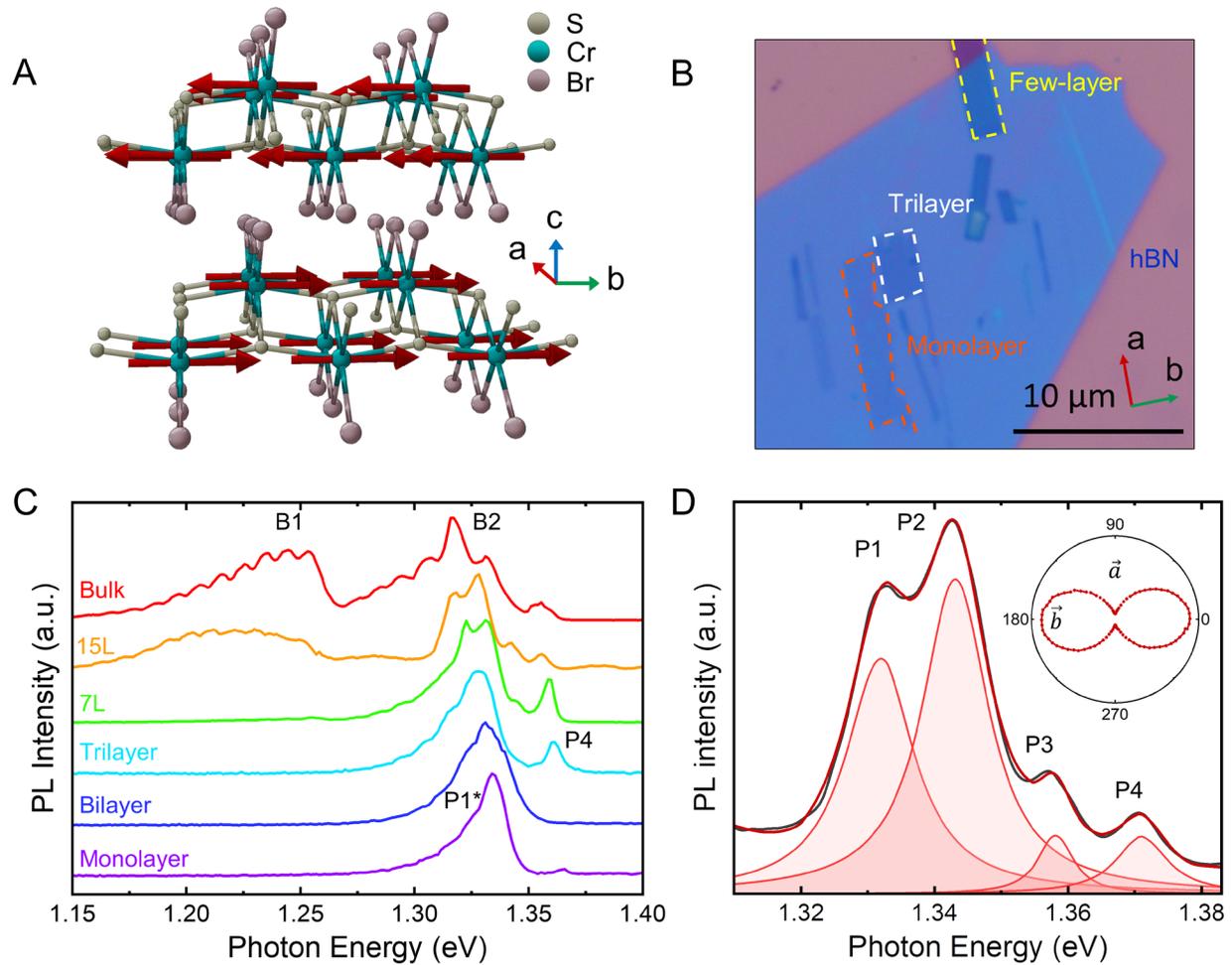

**Figure 1.** (A) Crystal structure of the layered magnetic semiconductor CrSBr. (B) Optical microscope image of different exfoliated CrSBr flakes over a silica on silicon substrate. (C) PL dependence on the thickness of different flakes at 4 K. (D) Lorentzian deconvolution for the emission at 4 K of the 15-layer CrSBr (15L). A polarization diagram is inserted in the figure to show that the emission is polarized along the $\vec{b}$ axis.

On the other hand, the emission of B2 appears at higher energies (in the range of 1.32-1.38 eV). These peaks have been ascribed to excitonic emission by several authors.[20,24,25]. For monolayer and bilayer, B2 appears as a single asymmetric peak at 1.332 eV and 1.330 eV, respectively (namely P1*). P1* broadens and splits into various peaks (at least three, namely P1, P2, and P3) in thicker flakes. But another peak, namely P4, is clearly seen at higher energies for trilayer and thicker flakes, including the bulk (1.361 eV for the trilayer). In the monolayer and bilayer, the contribution of P4 is negligible. A plot of the PL for a 15-layer crystal in which these peaks are deconvoluted and the polarization of the emission along the easy magnetic axis (*b*) is demonstrated is displayed in Figure 1D.

**Table 1. Energy splitting in CrSBr flakes.**

|  |  | This work | | Wilson et al. Ref. 20 | | Klein et al. Ref. 24 | | Cenker et al. Ref. 25 | |
|---|---|---|---|---|---|---|---|---|---|
|  | Peak | Energy (meV) | Linewidth (meV) | Energy (meV) | Linewidth (meV) | Energy (meV) | Linewidth (meV) | Energy (meV) | Linewidth (meV) |
| **Monolayer** | P1* | 1334 | 17 | 1320 | 14 | - | - | - | - |
| **Bilayer** | P1* | 1332 | 24 | 1320 | 24 | - | - | - | - |
| **Trilayer** | P1* | 1328 | 23 | 1330 | 16 | 1340 | 18 | - | - |
|  | P4 | 1361 | 6 | 1360 | 4 | 1367 | 7 | - | - |
| **Few-Layer (7L) †** | P1 | 1323 | 12 | 1325 | 5 | - | - | - | - |
|  | P2 | 1331 | 9 | 1340 | 9 | - | - | - | - |
|  | P3 | 1344 | 2 | - | - | - | - | - | - |
|  | P4 | 1359 | 4 | 1360 | 6 | - | - | - | - |
| **Multilayer (15L) ††** | P1 | 1318 | 12 | - | - | - | - | 1325 | 7 |
|  | P2 | 1328 | 9 | - | - | - | - | 1337 | 6 |
|  | P3 | 1342 | 7 | - | - | - | - | 1352 | 6 |

| | | | | | | | | |
|---|---|---|---|---|---|---|---|---|
| | P4 | 1356 | 4 | - | - | - | - | 1366 | 4 |
| Bulk (>50L) | P1 | 1307 | 12 | 1320 | 9 | - | - | - | - |
| | P2 | 1317 | 9 | 1328 | 8 | - | - | - | - |
| | P3 | 1332 | - | 1340 | 7 | - | - | - | - |
| | P4 | 1355 | 4 | - | - | 1366 | 2 | - | - |

†The few-layer of this work is composed by seven layers (7L) while in Ref. 16 it is a 4L.

†† The multilayer of this work is a 15L while in Ref. 20 authors estimated 20 nm (i.e. 17L).

As far as P1* is concerned, the shape of this peak is maintained almost unchanged until the trilayer with respect to the monolayer, although red-shifted (6 meV from monolayer to trilayer). We attribute this redshift to the variation in the vertical quantum confinement, as is typically seen in semiconductor nanostructures (along the vertical axis). [26–28] This matches the observations reported in previous works, see Table 1, which have been attributed to the strong interlayer electronic coupling in this material.[20] By studying the thermal and magnetic field dependence of the PL properties, we will demonstrate that the nature of P4 is very different from that of P1-P3, as this exhibits a very different temperature and magnetic field dependence.

TEMPERATURE EFFECT

We focus on the 15L CrSBr case (15 layers), as this flake exhibits the clearest signature of charge transfer between excitonic species (Figures 2A and 2B). The temperature evolution of other flakes is included as supporting information (see Supporting Information Section SI1). At 4 K the PL is dominated by P1-P3 with a minor contribution of P4. In the temperature range 15-50 K the emission of P4 shows a clear increase (see Figures 2C and 2D). At higher temperatures, the splitting of the band is hindered by the spectral broadening and the contribution of the different peaks, obtained by Lorentzian fittings (see details in Supporting Information Section SI3). At first sight, the increasing contribution of P4 at expenses to the other peaks might reminisce about the exciton dynamics in TMDCs, where the contribution of trion and the defect-bound excitons feeds the neutral exciton emission. However, in contrast to the behavior reported in TMDC, this trend occurs at lower temperature, and it is reversed with the transfer from P4 to P2 at higher temperature.

By plotting the integrated intensity of PL as a function of the inverse of the temperature we can extract information required to discuss about the mechanism involved in the recombination dynamics. These Arrhenius plots are reported in Fig. 3 and parametrized in terms of activation energies ($E_i$), scattering rates ($G_i$) and the fraction of P4 moving to P2 and or vice versa ($A_3$ and $A_5$, respectively) (see Supporting Information Section SI2 for further details). The results are summed up in Table 2.

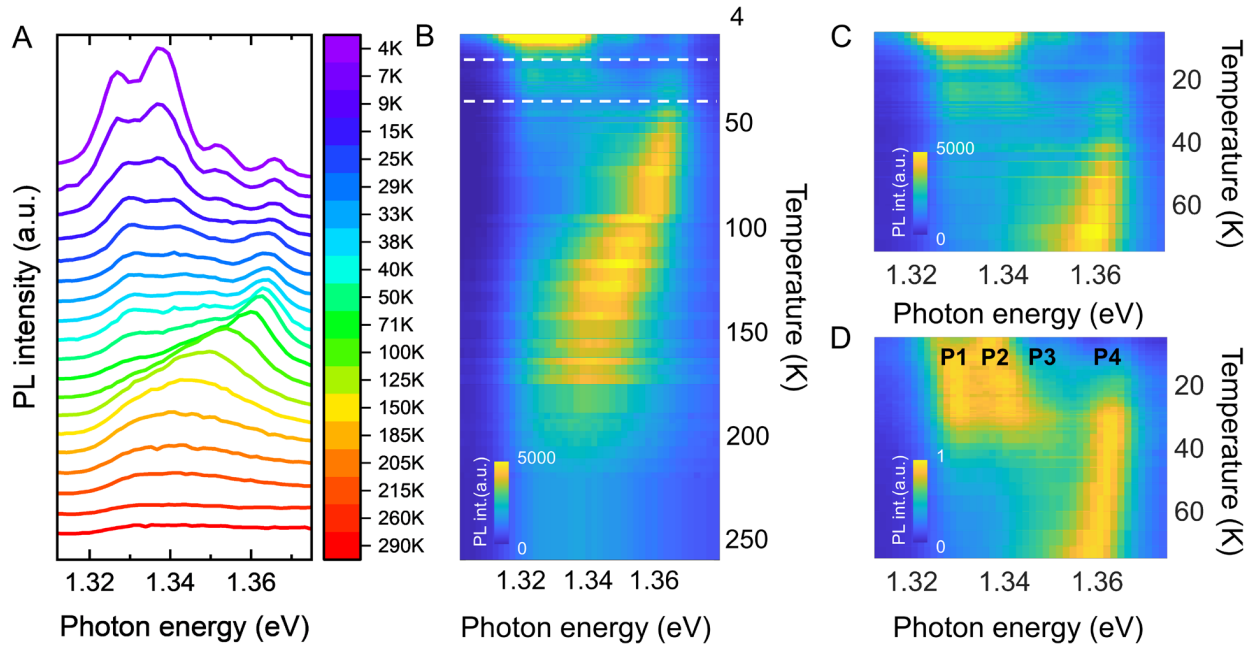

**Figure 2.** (A) PL dependence on the temperature for a multilayer (15L) CrSBr crystal. (B) Color map composed by the spectra in (A). The region of interest is zoomed out in (C) and (D). In (D) the PL intensity has been normalized to the maximum of each line to highlight the contrast.

**Table 1.** Parameters of the activation mechanism for P1, P2, P3, and P4.

| Mech. | P1 | | P2 | | P3 | | P4 | |
|---|---|---|---|---|---|---|---|---|
| | Rate | $E_a$ (meV) | Rate | $E_a$ (meV) | Rate | $E_a$ (meV) | Rate | $E_a$ (meV) |
| 1 | 2.6 | 1.2 | 4.2 | 1.5 | 10 | 4.1 | - | - |
| 2 | 26 | 14 | - | - | - | - | - | - |
| 3 | - | - | 0.01 | 35 ($A_3$=2.6) | - | - | 120.7 | 39 |
| 4 | - | - | 2090 | 72 | - | - | - | - |
| 5 | - | | | | - | - | 0.14 | 3.7 ($A_5$=2.1) |

The activation energies of P1-P3 (i.e 1.2, 1.5 and 4.1 eV respectively) are characteristic non-radiative mechanisms related with the ionization of vacancies and impurities, very common in semiconductor quantum wells.[29] The behavior of P2 and P4 is much more interesting as their contribution to the PL is correlated. Thus, while P2 shows a minimum in PL intensity at 40-50 K, P4 exhibits a sharp maximum in this temperature range. Simultaneously to the quenching of P2, the increase in the PL intensity of P4 upon warming up is due to a feeding mechanism involving a similar activation energy [E5(P4) = 3.7 meV]. In contrast, at higher temperatures the intensity of P4 quenches in favor of P2. Again, the activation energies of the involved mechanisms are similar [E3(P2) = 35 meV; E3(P4) = 39 meV].

The above discussion clearly evidences the different nature of P1, P2 and P3 with respect to P4. Thus, the marked increase in the ionization of donor impurities upon warming up lead to a quenching in the PL of P1, P2 and P3, as expected for excitons. However, the increase of uncorrelated electrons in the conduction band will boost the emission of P4, as expected for charged excitons.

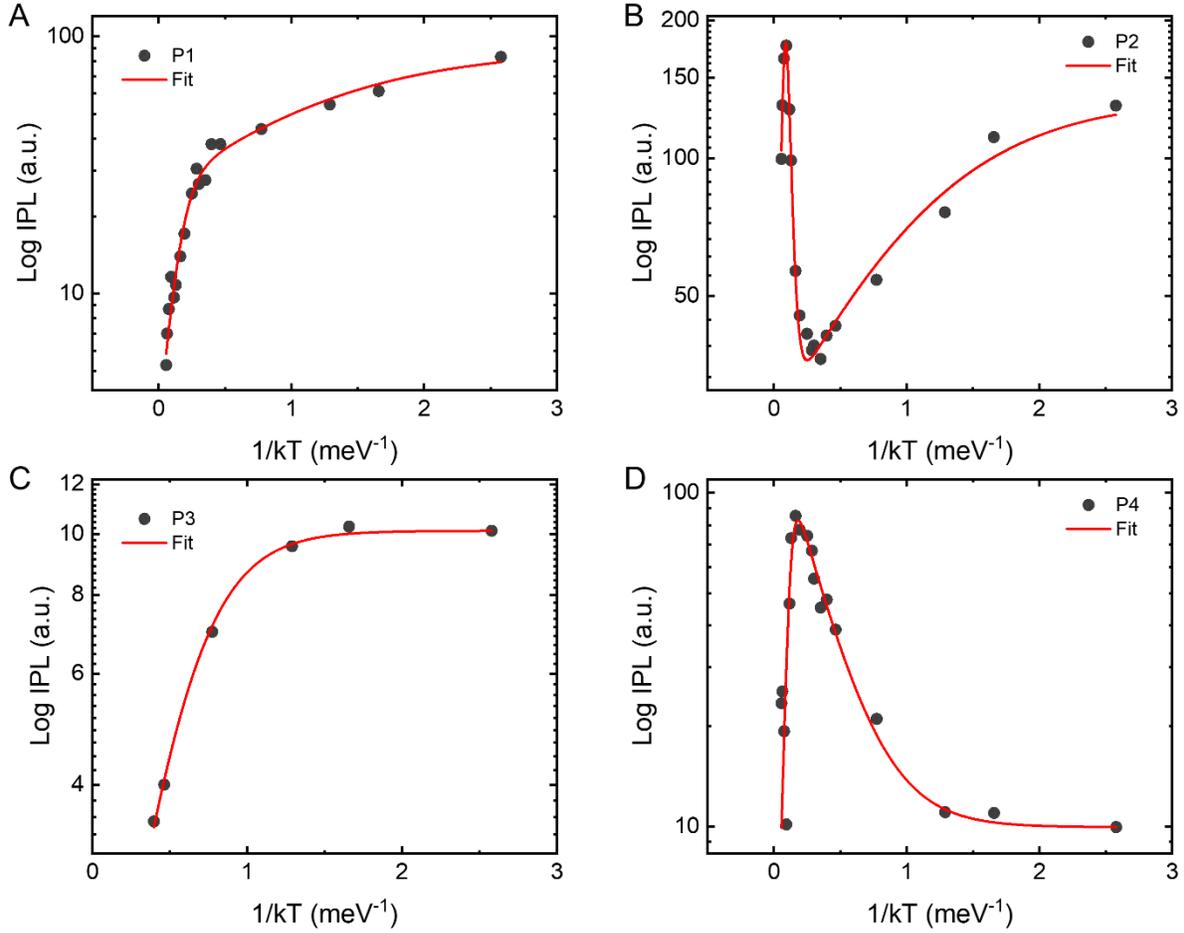

**Figure 3.** Arrhenius plots for the different peak emission (A)-(D) corresponding to P1-P4. Values of the integrated area are obtained by Lorentzian deconvolution of the experimental spectra of B2 and then fitted with the corresponding expression in Supporting Information Section 2. The values of the fit are summarized in Table 2.

EFFECT OF THE MAGNETIC FIELD

The different nature of P4 (with respect to P1-P3) is further supported by the magnetic field dependence of this peak. In Figure 4 we plot the effect of the magnetic field applied along the easy *b*-axis over the PL properties of CrSBr layers of two different thickness (3 and 15 layers) at 4 K. The field-induced transition from AP to P alignment of the ferromagnetic monolayers is accompanied by a redshift in the peaks P1-P3, in agreement with previous reports.[20,25] In the trilayer case, this field-induced magnetic switching ($B_{FLIP}$) occurs at *ca.* 0.2 T (Figure 4A and B), as expected according to magneto-transport measurements,[14] while for thicker layers (15L) $B_{FLIP}$ increases to 0.45 T (Figure 4C and D). P4 is only observed in the AP spin configuration, i.e., below the switching field, being fully quenched in the field-induced P configuration [without exhibiting any energy shift (Figure 4B and D)]. It is worth noting that in the thicker layers the contribution of B1, previously ascribed to the emission of donor vacancies in bulk CrSBr,[17] is also spin dependent. Indeed, a clear intensity decrease is observed for the P configuration coinciding with the quenching of P4.

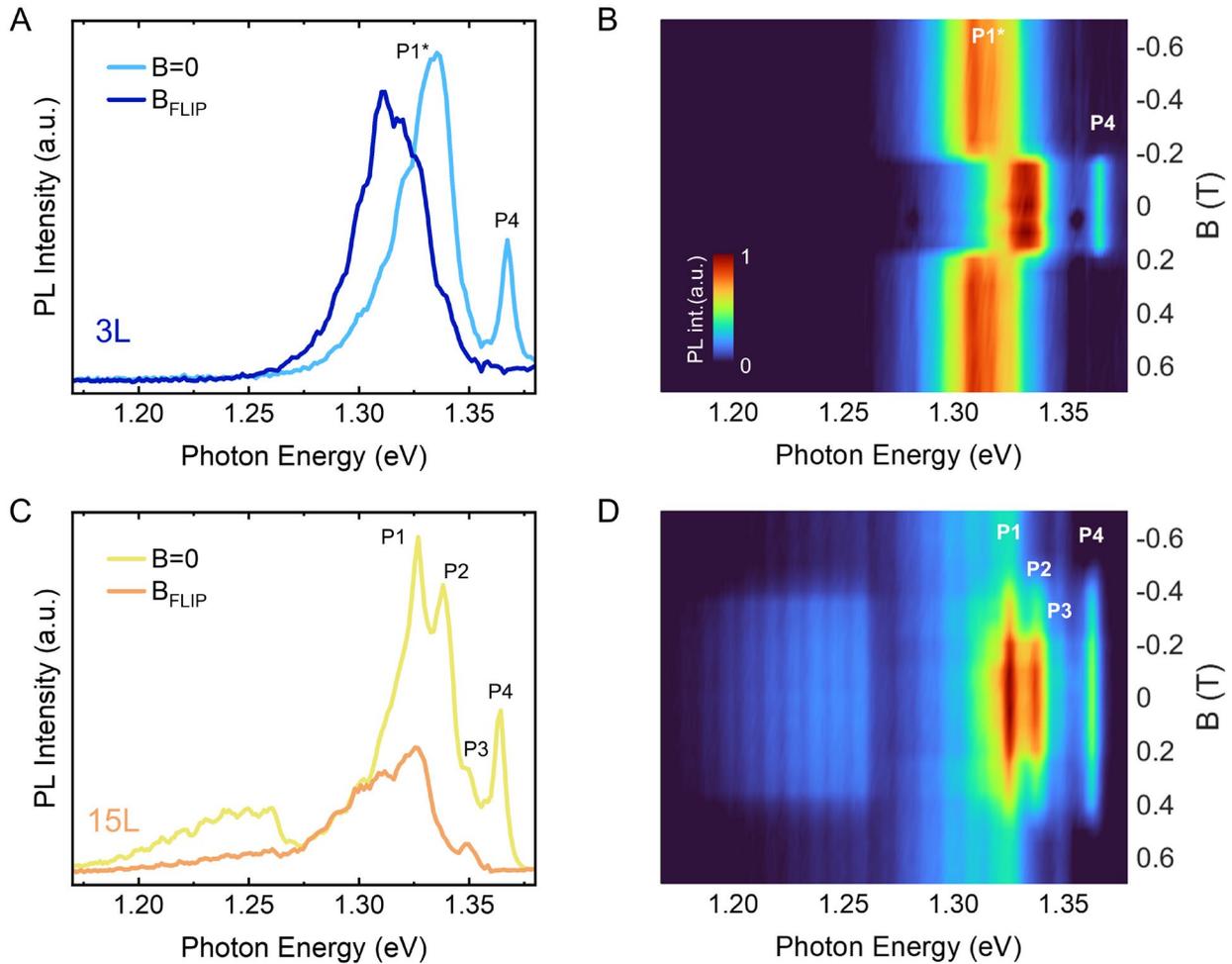

**Figure 4. PL dependence on the external magnetic field along the easy axis. (A) and (C) show the PL at 4 K, for the trilayer and 15L, respectively, exposed to two different magnetic fields,**

i.e., 0 and 0.6 Tesla. (B) and (D) PL vs the external magnetic field for the trilayer and multilayer, respectively.

To explain the above observations, we summarize in Scheme 1 the magnetic field impact on the excitonic properties, underlying the effects of the electronic delocalization. In the AP configuration the electron is localized within a given layer (scheme 1A and C), while it becomes delocalized over different layers in the field-induced P configuration (scheme 1B and D).[20] Thus, the probability of recombination of the electron-hole pair is dramatically reduced when the electron and hole are located in different layers (inter-layer case in the AP configuration, Scheme 1C). This explains why the emission for the monolayer is field-independent, while a redshift is observed in the few-layer cases.

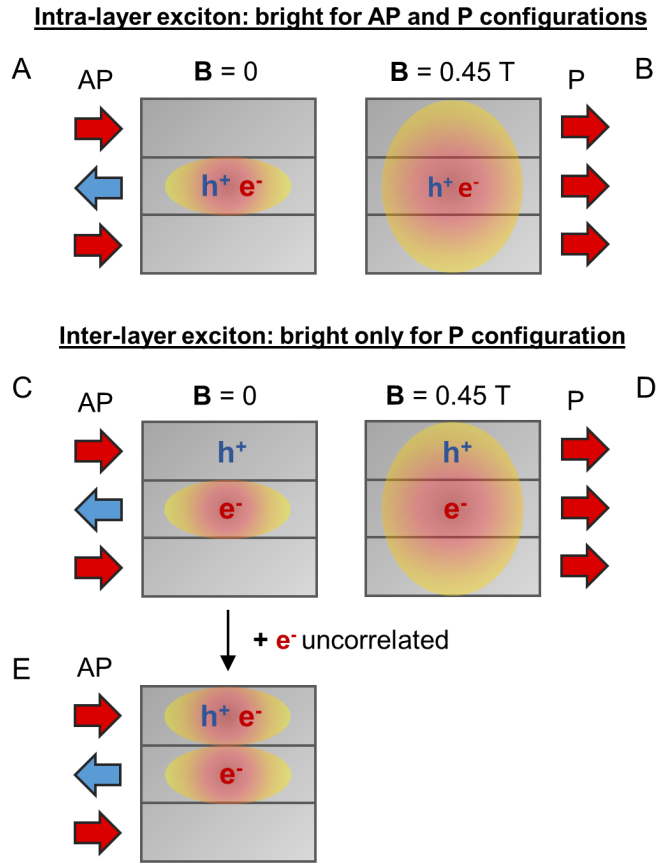

**Scheme 1. Intra-layer vs. inter-layer excitons and influence of the magnetic configuration**
In the AP configuration, the electron wavefunction is localized in a layer; hence, an intra-layer electron-hole pair recombination is required for the emission of light (case A), while the probability for an inter-layer electron-hole pair recombination is lower (case C). In the field-induced P configuration, the electron wavefunction is delocalized over different layers; hence, an interlayer electron-hole pair recombination is also possible (case D). This leads to a redshift of the exciton emission. In case C (inter-layer exciton in AP configuration), emission is still possible if an extra electron is injected in the layer where the hole is located, as it can activate a trion recombination channel (case E). This trion recombination channel is deactivated in the P configuration since the recombination of inter-layer excitons is preferred (case D).

Summing up, we have demonstrated that the behaviour of P4 differs from the peaks commonly ascribed to excitons because: it presents a different recombination dynamic, its emission is boosted by ionization of impurities, and it is quenched in the P spin configuration. Those properties will be essential for the further discussion about the nature of P4.

Firstly, we compare P4 with the emission peak observed at the same energy by other authors. Particularly, Klein and co-workers identified a one-dimensional exciton (namely 1S exciton) occurring at an emission energy close to our P4. However, ascribing P4 to the 1S exciton cannot explain its dependence on the magnetic field or its temperature evolution. In those experiments, 1S dominates the emission in the monolayer limit while P4 is negligible in our monolayers. Moreover, the temperature evolution of 1S is very different from that observed for P4. Then, it is obvious that P4 corresponds to a different kind of emission.

Secondly, we compare the magnetic field dependence of P4 with the rest of peaks in B2. As described above, Wilson et al. foresee (and demonstrate experimentally) a strong electron confinement in the antiparallel spin configuration which reduces the oscillator strength of "interlayer excitons" (i.e. excitons with the hole and the electron in different layers). In the case of P4, we observe a clear quenching coinciding with the redshift of the other peaks due to delocalization. Hence, the emission of P4 must be related to the intra-layer electron localization in the AP spin configuration.

Thirdly, we compare the impact of the ionization of impurities on the emission of the different peaks. During the preparation of this work, it has been demonstrated that the magnetic impurities (most probably Br and S vacancies) play an important role in the magnetic and magneto-resistive response of few-layer CrSBr. It is also stablished that the connection between the optical and the magnetic properties is mediated by those impurities. Our results verify this hypothesis, as we are presenting clear evidence that the ionization of magnetic (donor) impurities is the link between magnetism and optical properties. Importantly, the onset of P4 with the ionization of impurities suggests that this peak could be a charged exciton, and therefore, it could be tentatively ascribed to a negative trion: a trion formed with the injection of a free electron into "dark interlayer" exciton, as described in the Scheme (from 1C 1E).

Notice that the observation of an antibonding trion is not usual in 2D systems. In general, negative trions tend to be bonding as their electrons are paired (one spin up and one spin down). In contrast, the electrons in the conduction band of CrSBr would be unparied given the atypical electronic structure of this material. In this situation the exchange interaction could positive (electron repulsion) generating an antibonding specie. Moreover, the emission of antibonding species is only possible under specific electron confinement. In the case of CrSBr, the generation of "dark interlayer" excitons with a very low oscillator strength could be seed for the generation of antibonding charged species. A rigorous assignment would require of further theoretical and experimental developments **out of the scope of this work which focus on use of optical measurements for monitoring the magnetic states of CrSBr**.

**CONCLUSIONS**

To conclude, we want to highlight that our results provide an excellent tool to obtain magnetic information of CrSBr flakes by means of optical techniques. Thus, we have identified the different mechanisms explaining the exciton dynamics in CrSBr down to the monolayer. Our measurements reveal the existence of a particular transition (P4) which contribution is boosted by the ionization of donor magnetic vacancies (S or Br). This explains the connection between optical and magnetic properties since both are bound to these kinds of defects. In addition, we observed that P4 is very sensitive to external magnetic fields, as it is only observed in the AP magnetic configuration. This can enable the optical reading of magnetic states in spintronic and optoelectronic devices by means of optical read out. Finally, we pointed out the quenching of the neutral exciton emission in favour of P4 is observed in the temperature range 30-40K, coinciding with the hidden order transition.

**METHODS**

The thickness of the studied flakes was identified with an optical microscope by a calibrated contrast comparison method.[14] Micro-photoluminescence measurements were carried out on a µPL confocal system coupled to an attoDRY1000 cryostat and provided with a set of superconducting magnets. Inside the cryostat, the samples were placed on top of a piezoelectric three-axial stage for accurate positioning of single flakes. A laser diode at 520 nm wavelength with a power control driver was used as excitation source. The diode emission was coupled to a monomode optical fiber to obtain a diffraction limited spot with a power ranging from 0 to 100 µW. Then, backscattering was collected by means of a monomode optical fiber filtering the laser to obtain the µPL signal which was analyzed with cooled silicon back-thinned CCD attached to a spectrometer.

ACKNOWLEDGMENT

The authors acknowledge the financial support from the European Union (ERC AdG Mol-2D 788222 and FET OPEN SINFONIA 964396), the Spanish MCIN (2D-HETEROS PID2020-117152RB-100, co-financed by FEDER, and Excellence Unit "María de Maeztu" CEX2019-000919-M), the Generalitat Valenciana (PROMETEO Program, PO FEDER Program IDIFEDER/2018/061, a Ph.D fellowship to C.B.-C., a postdoctoral fellow APOSTD-CIAPOS2021/215 to S.M.-V. and the grant CIDEGENT/2018/005 to J.C.-F.). We thank Á. López-Muñoz for their constant technical support and helpful discussions.